# Magnon squeezing in an antiferromagnet: reducing the spin noise below the standard quantum limit

Jimin Zhao[*‡], Andrea V. Bragas[*‡], David J. Lockwood[†] & Roberto Merlin[*]

[*] *FOCUS Center and Department of Physics, The University of Michigan, Ann Arbor, Michigan 48109-1120, USA*

[†] *Institute for Microstructural Sciences, National Research Council, Ottawa K1A 0R6, Canada*

[‡] *These authors contributed equally to the work*

………………………………………………………………………………………………………..

At absolute zero temperature, thermal noise vanishes when a physical system is in its ground state, but quantum noise remains as a fundamental limit to the accuracy of experimental measurements. Such a limitation, however, can be mitigated by the formation of squeezed states. Quantum mechanically, a squeezed state is a time-varying superposition of states for which the noise of a *particular* observable is reduced below that of the ground state at certain times. Quantum squeezing has been achieved for a variety of systems, including the electromagnetic field, atomic vibrations in solids and molecules, and atomic spins, but not so far for magnetic systems. Here we report on an experimental demonstration of spin wave (i.e., magnon) squeezing. Our method uses femtosecond optical pulses to generate correlations involving pairs of magnons in an antiferromagnetic insulator, $MnF_2$. These correlations lead to quantum squeezing in which the fluctuations of the magnetization of a crystallographic unit cell vary periodically in time and are reduced below that of the ground state quantum noise. The mechanism responsible for this squeezing is stimulated second order Raman scattering by magnon pairs. Such squeezed states have important ramifications in the emerging fields of spintronics and quantum computing involving magnetic spin states or the spin-orbit coupling mechanism.

The fact that noise is inherent to quantum systems has been known since Heisenberg postulated the uncertainty principle in 1927.[1] Remarkably, despite this fact, the noise of a *given* quantum observable can in theory be eliminated entirely through quantum squeezing.[2] Quantum squeezing refers to a state whereby noise in one variable is reduced at the expense of enhancing the noise of its conjugate variable. The terminology originates from quantum optics and it is usually applied to a system of non-interacting bosons. Squeezed states have only been achieved fairly recently and were first demonstrated for the electromagnetic field (photons) in 1985.[3] Nonlinear interactions of light with passive and active atomic media have been successfully used to generate squeezed photons, opening possibilities for essentially noiseless optical communications and precision measurements.[4,5] The electromagnetic field is now not the only system that has been squeezed, and thermal squeezing is also possible for classical objects.[6] In the past few years, squeezed states of vibrational degrees of freedom have been experimentally demonstrated for molecules[7] and solids[8,9] and, more recently, squeezed atomic spin states have also been achieved.[10-14] Because spin squeezing is closely related to quantum entanglement, squeezed states hold promise for application sin quantum computing.[15]

Impulsive stimulated second-order Raman scattering (RS) has been shown to be a practical means to generate squeezed states for phonons in solids.[8,9,16] Second-order Raman coupling between the incident light and lattice vibrations is proportional to the square of the phonon amplitude. This interaction gives rise to squeezing, because it represents a change in the frequency of the harmonic oscillator for the duration of the pulse and, as such, it is a parametric perturbation. Here, we concern ourselves with noise in a magnetically-ordered solid for which the low-lying spin excitations are bosons known simply as magnons. As for phonons, we achieve magnon squeezing by means of impulsive second-order RS. Magnon squeezing manifests itself in the reduction of the noise of the local crystallographic cell magnetization.

Manganese difloride, MnF$_2$, crystallizes in the rutile structure (of $D_{4h}$ symmetry) and becomes antiferromagnetic below the Néel temperature $T_N = 68$ K.[17] In the ordered state, there are two sublattices, α and β, in which the Mn$^{2+}$ spins align "up" and "down" along the four-fold axis of symmetry [001]. Ignoring the weaker intrasublattice exchange and the magnetic anisotropy, the spin dynamics of MnF$_2$ is well described by the one-parameter Hamiltonian

$$H_0 = J \sum_{\langle u,v \rangle} \mathbf{S}_{u,\alpha} \cdot \mathbf{S}_{v,\beta} \quad . \quad (1)$$

Here, $J$ is dominant intersublattice exchange constant associated with the interaction between a given Mn$^{2+}$ ion and its eight next-nearest neighbors and the sum runs over pairs of next-nearest neighbors.[17] In the harmonic approximation, $H_0$ becomes

$$H_0 \approx \sum_{\mathbf{q}} \hbar \Omega_{\mathbf{q}} (a^+_{\uparrow \mathbf{q}} a_{\uparrow \mathbf{q}} + a^+_{\downarrow \mathbf{q}} a_{\downarrow \mathbf{q}}) \quad (2)$$

where $a^+_{\uparrow \mathbf{q}}$ and $a^+_{\downarrow \mathbf{q}}$ ($a_{\uparrow \mathbf{q}}$ and $a_{\downarrow \mathbf{q}}$) are creation (annihilation) operators for magnons of wave vector $\mathbf{q}$ and frequency $\Omega_{\mathbf{q}}$ belonging to the two degenerate branches labeled ↑ (up) and ↓ (down).[18-19] We recall that the degeneracy can be lifted by an external magnetic field and that, for wave vectors near the edge of the Brillouin zone, the excitations generated by $a^+_{\uparrow \mathbf{q}}$ ($a^+_{\downarrow \mathbf{q}}$) propagate mainly on the α (β) sublattice (a magnon of arbitrary $\mathbf{q}$ generally perturbs both sublattices).[20]

Phenomenologically, the Raman coupling between magnons and light can be expanded in powers of the spins of the magnetic ions. In MnF$_2$, and most antiferromagnets, the first-order contribution relying on spin-orbit coupling is far less important than the second-order term which results from excited-state exchange and leads to RS by magnon pairs. Let $\mathbf{E}(\mathbf{r},t) = E(e_x, e_y, e_z)$ be the position-and time-dependent electric field. From symmetry considerations, the two-magnon interaction relevant to stimulated RS can be written as[18,19]

$$V = \frac{E^2}{2} \sum_{\langle u,v \rangle} \Xi(v,w)(S^+_{u,\alpha}S^-_{v,\beta} + S^-_{u,\alpha}S^+_{v,\beta} + \gamma S^z_{u,\alpha}S^z_{v,\beta}) \quad (3)$$

where $S_\pm = S_x \pm iS_y$ and

$$\Xi(v,w) = \kappa_1(e_x e_x + e_y e_y) + \kappa_2 e_z e_z + 2\kappa_3[e_x e_y \operatorname{sgn}\sigma_x \operatorname{sgn}\sigma_y] \\ + 2\kappa_4[e_y e_z \operatorname{sgn}\sigma_y \operatorname{sgn}\sigma_z + e_x e_z \operatorname{sgn}\sigma_x \operatorname{sgn}\sigma_z] \quad . \quad (4)$$

Here $\gamma$ is an anisotropy parameter, $\boldsymbol{\sigma} = (\sigma_x, \sigma_y, \sigma_z)$ is a vector connecting a given ion with its next-nearest neighbors in the opposite sublattice, and $\kappa_m$ ($m = 1$-$4$) are magneto-optic constants associated with the Raman tensors of symmetry $A_{1g}$ ($\kappa_1$ and $\kappa_2$), $B_{2g}$ ($\kappa_3$), and $E_g$ ($\kappa_4$). The relative magnitudes of the latter can be inferred from spontaneous Raman data which give $|\kappa_1/\kappa_3| = 0.14$, $|\kappa_2/\kappa_3| = 0.32$ and $|\kappa_4/\kappa_3| = 0.66$.[19] In our experiments, the only symmetry that matters is $B_{2g}$.

It is straightforward to express $V$ in terms of magnon variables. For our purposes, however, it is sufficient to consider only zone-edge excitations for the magnon density of states is strongly peaked at van Hove singularities close to the Brillouin zone boundary. Writing $V = \sum_\mathbf{q} V_\mathbf{q}$, we obtain for $\mathbf{q}$ near the edge

$$V_\mathbf{q} \approx SE^2 \Delta_\mathbf{q} (a^+_{\downarrow\mathbf{q}} a^+_{\uparrow-\mathbf{q}} + a_{\downarrow\mathbf{q}} a_{\uparrow-\mathbf{q}}) \quad (5)$$

where $S = 5/2$ is the spin of $Mn^{2+}$ and $\Delta_\mathbf{q}$ is a weighting factor that generally depends on the field polarization and the appropriate combination of magneto-optic coefficients. For the $B_{2g}$ term, $\Delta_\mathbf{q} = -8\kappa_3 \sin(q_x a/2)\sin(q_y a/2)\cos(q_z c/2)$ where $a$ and $c$ are the lattice constants.[18,19] This geometry favors the M-point of the Brillouin zone and gives the largest contribution to two-magnon RS.[18,19]

- 4 -

The coupling defined by *V* describes a time-dependent pair-wise interaction between magnons. To describe the effect of a laser pulse on the magnetic system, we consider the impulsive limit, i.e., an optical pulse of width $\tau \ll \Omega^{-1}$ and, for simplicity, we take the speed of light $c_L \to \infty$. Then, the field can be treated as an instantaneous and position-independent perturbation $E^2 \approx (4\pi I / n_R c_L)\delta(t)$ where $n_R$ is the refractive index and *I* is the integrated intensity of the pulse. Let $\Psi_0$ be the wavefunction of the whole crystal just before the pulse strikes. A simple integration of the Schrödinger equation gives, for $t > 0$,[8]

$$\Psi = e^{iH_0 t/\hbar} \exp\left[-i\sum_{\mathbf{q}} \frac{4\pi SI}{n_R c_L \hbar} \Delta_{\mathbf{q}} (a^+_{\downarrow \mathbf{q}} a^+_{\uparrow -\mathbf{q}} + a_{\downarrow \mathbf{q}} a_{\uparrow -\mathbf{q}})\right]\Psi_0 \quad . \quad (7)$$

At zero temperature, $\Psi_0$ is the magnetic ground state and this expression becomes identical to that describing two-mode squeezed states.[2] After some algebra, we obtain to lowest order in the intensity of the pulse

$$\langle \Psi | (a^+_{\downarrow \mathbf{q}} a^+_{\uparrow -\mathbf{q}} + a_{\downarrow \mathbf{q}} a_{\uparrow -\mathbf{q}}) | \Psi \rangle \approx -\frac{8\pi SI}{n_R c_L \hbar} \Delta_{\mathbf{q}} \sin(2\Omega_{\mathbf{q}} t) \quad (8)$$

Hence, equation (7) represents a state in which there exists a time-varying correlation between magnons of opposite wave vectors that belong to the two degenerate branches.

Data were obtained from a $5.5 \times 5.8 \times 6.0 \text{ mm}^3$ single crystal of $MnF_2$. Time-domain measurements were performed at 4 K using a standard pump-probe setup in a transmission geometry that allows only $B_{2g}$ excitations (see Methods). Light penetrated the crystal along the [001] direction. We used 50 fs pulses generated by an optical parametric amplifier at a repetition rate of 250 kHz and 534 nm central wavelength, which were focused onto a common 30-μm-diameter spot. The average power for the pump and probe pulses was, respectively, 13 and 3 mW. The stronger pump pulse drives the crystal into the time-varying squeezed state [equation (7)], and the concomitant time-varying refractive index scatters the weaker probe pulse that follows behind. The signal of interest is the transmitted intensity of the probe beam as a function of the time delay between the two pulses. Using equation (8) and well-known results for coherent phonons,[21] we obtain the following expression for the differential transmission

$$\Delta T/T \approx \frac{\ell I}{\hbar N v_C} \left(\frac{8\pi S}{n_R c_L}\right)^2 \sum_q \Omega_q \Delta_\mathbf{q}^2 \cos(2\Omega_q t) \quad (9)$$

which is valid for probe pulses of negligible width. Here $\ell$ is the length of the sample, $N$ is the number of unit cells and $v_C$ is the cell volume.

The time-domain data of Fig. 1 show well-resolved oscillations. After removal of the so-called coherent artifact at $t = 0$, we used linear prediction methods[22] to determine the number of oscillators and their parameters. As shown in the Fourier transform spectrum of the inset, this procedure gives three modes. Their positions, at ~100, 347 and 481 cm$^{-1}$, are in excellent agreement with those of, respectively, the two-magnon feature and the Raman-allowed phonons of symmetries $A_{1g}$ and $B_{2g}$.[23] Based on this result and the previous discussion, the ~ 100 cm$^{-1}$ oscillations are ascribed to a two-magnon squeezed state whereas the other two features are assigned to coherent phonons.[23] The observation of the $A_{1g}$ mode is attributed to a polarization leakage in the experiment. While $A_{1g}$ excitations are not nominally allowed in our configuration, the Raman cross section for the 347 cm$^{-1}$ phonon is so large that it cannot be entirely suppressed by our method (see Methods).

The two-magnon oscillations are reproduced in Fig. 2(a) after numerical subtraction of the signal due to the pair of phonons. Figs. 2(b) and (c) compare the Fourier transform of the time-domain data and the conventional two-magnon Raman spectrum. In Fig. 2(d), we show the magnon dispersion curve,[17] which supports our contention that the pump-probe signal is dominated by magnon pairs at the zone boundary. The magnetic oscillations do not show the phase predicted by equation (9) at zero time delay. We believe that this discrepancy reflects simply the fact that equation (3) and, therefore, equation (9) apply to transparent materials whereas the wavelength of our laser falls within a broad absorption band of MnF$_2$.[24] In such a case, it is well-known that the phase of the coherent oscillations can have arbitrary values.[26] Further support for this interpretation is the observation that the two-magnon signal does not vanish in integrated transmission measurements reflecting the fact that the sample behaves as a frequency-dependent filter.[21]

The above arguments support our claim that the coherence responsible for the oscillations of Fig. 2 is that of a magnon squeezed state, but we have not yet provided a physical interpretation of such a state. Let $\mathbf{m}_l$ be the local magnetization operator. In the following we show that magnon squeezing is tantamount to time-dependent fluctuations of the local magnetization noise $\Delta m_l = \sqrt{\langle \mathbf{m}_l^2 \rangle}$ (note, however, that $\langle \mathbf{m} \rangle = 0$). The proof is relatively simple. Writing $\mathbf{m}_l = (g\mu_B/\hbar)(\mathbf{S}_{l,\alpha} + \mathbf{S}_{l,\beta})$ where the spins belong to the same unit cell, $g$ is the Landé factor and $\mu_B$ is the Bohr magneton, we have

$$\frac{1}{2}\left(\frac{\hbar}{g\mu_B}\right)^2 \sum_l \langle \mathbf{m}_l^2 \rangle = NS(S+1) + \sum_{\mathbf{q}} \langle \mathbf{S}_\alpha(\mathbf{q}) \cdot \mathbf{S}_\beta(-\mathbf{q}) \rangle \qquad (10)$$

where $N$ is the total number of unit cells and $\mathbf{S}_{\alpha/\beta}(\mathbf{q}) = \sum \mathbf{S}_{l,\alpha/\beta} \exp(i\mathbf{q}\mathbf{r}_{l,\alpha/\beta})/N^{1/2}$ (the sum is over all sites of the corresponding sublattice). Since $\Delta m_l$ does not depend on the particular unit cell we omit the sub-index in the following. From equations (7) and (10), we obtain in the harmonic approximation and for $I \to 0$

$$\Delta m(t) \approx (2S)^{1/2} \left(\frac{g\mu_B}{\hbar}\right) \left[1 + \frac{4\pi IS}{Nn_R c_L \hbar} \sum_q \Delta_\mathbf{q} \sin(2\Omega_q t)\right]. \quad (11)$$

If we compare this with equation (9) for $\Delta T/T$, it is clear that the amplitude of the coherent oscillations is proportional to that of the local magnetization noise. In essence, the noise is being controlled by the laser-induced refractive index modulation. Since both equations (9) and (11) are dominated by contributions from magnons near the M-point of the Brillouin zone, of frequency $\Omega_M$, we obtain approximately

$$\Delta m(t) \approx \Delta m(0) \left[1 + I\sqrt{\frac{v_C \xi}{4\hbar \Omega_M}} \sin(2\Omega_M t)\right] \quad (12)$$

where $\xi = \frac{1}{\ell I}\left.\frac{\Delta T}{T}\right|_{max}$ and $\Delta m(0) = (2S)^{1/2}\frac{g\mu_B}{\hbar}$ is the zero-point noise. A pictorial representation of a squeezed state is shown in Fig. 3. Instead of the local magnetization itself, we use the angle θ defined as $\cos\theta = \mathbf{S}_l^\alpha \cdot \mathbf{S}_l^\beta / S(S+1)$ to represent the noise. It is useful at this point to relate the magnetization noise to the coupling to light. A simple calculation shows that $V$ can be expressed as $V = E^2 \sum_\mathbf{q} \Theta(\mathbf{q}) \mathbf{S}_\alpha(\mathbf{q}) \cdot \mathbf{S}_\beta(-\mathbf{q})$. It follows that, except for the weighting factors $\Theta(\mathbf{q})$, light couples directly to fluctuations of the local magnetization. Given that contributions near the Brillouin zone boundary are dominant, we have approximately $V \approx E^2 \widetilde{\Theta} N (\hbar \Delta m / g\mu_B)^2 / 2$ where $\widetilde{\Theta}$ denotes an average value at the zone boundary. By evaluating (12) with the experimental values one can estimate the noise reduction. The thermal noise at 4K is approximately the product of $\Delta m(0)$ times twice the Bose factor at $\hbar\Omega_M$.[9] Under our experimental conditions the term multiplying $\sin(2\Omega_M t)$ in equation

(12) is $\sim 2 \times 10^{-5}$ which surpasses with confidence the thermal contribution to the noise which is $\sim 3 \times 10^{-8}$. Clearly, the noise level in the local magnetization has been reduced below the quantum limit.

## Methods

We used pump pulses polarized along the [110] direction. This geometry couples to modes of both $A_{1g}$ and $B_{2g}$ symmetry. The probe polarization was set at an angle of 45º with respect to the polarization of the pump. After going through the sample, the transmitted probe pulses were divided into two beams, one polarized parallel and the second one perpendicular to the pump polarization. These beams were sent to two separate detectors. It can be shown that the difference between the signals recorded by these detectors is equal to twice the differential transmission for the $B_{2g}$ geometry while the subtraction eliminates the isotropic $A_{1g}$ contribution.

**Acknowledgments** Work supported by the NSF under Grants No. PHY 0114336 and No. DMR 0072897, and by the AFOSR under contract F49620-00-1-0328 through the MURI program. One of us (AVB) acknowledges partial support from CONICET, Argentina.

**Competing financial interests** The authors declare that they have no competing financial interests.

**Correspondence** and requests for materials should be addressed to R. M. (merlin@umich.edu)

Figure Legends

**Figure 1 Time-domain data.** Relative differential transmission as a function of pump-probe time delay. The associated Fourier transform spectrum in the inset shows peaks due to two-magnon excitations (squeezed magnons) and coherent phonons.

**Figure 2 Two magnon results. a,** Pump-probe data showing two-magnon oscillations. **b,** Fourier transform spectrum. **c,** Raman spectrum recorded at 3 K using 140 mW of 514.5nm Ar laser light. **d,** Magnon dispersion for wave vectors in the [100] and [001] directions (to compare with the experiments, the frequency scale has been multiplied by a factor of two).

**Figure 3 Spin squeezing in MnF$_2$.** The arrows and cones attached to the manganese ions represent the spins and their angular fluctuations at zero temperature. The inset shows the time-dependence of the noise (represented by the angle $\theta$, which fluctuates about an expectation value of zero) and the square root of the expectation value of $\theta^2$, which is finite. The arrow indicates the time at which the laser pulse impinges on the solid, which via the two-magnon spin squeezing mechanism results in the oscillations in $(\langle\theta^2\rangle)^{1/2}$.

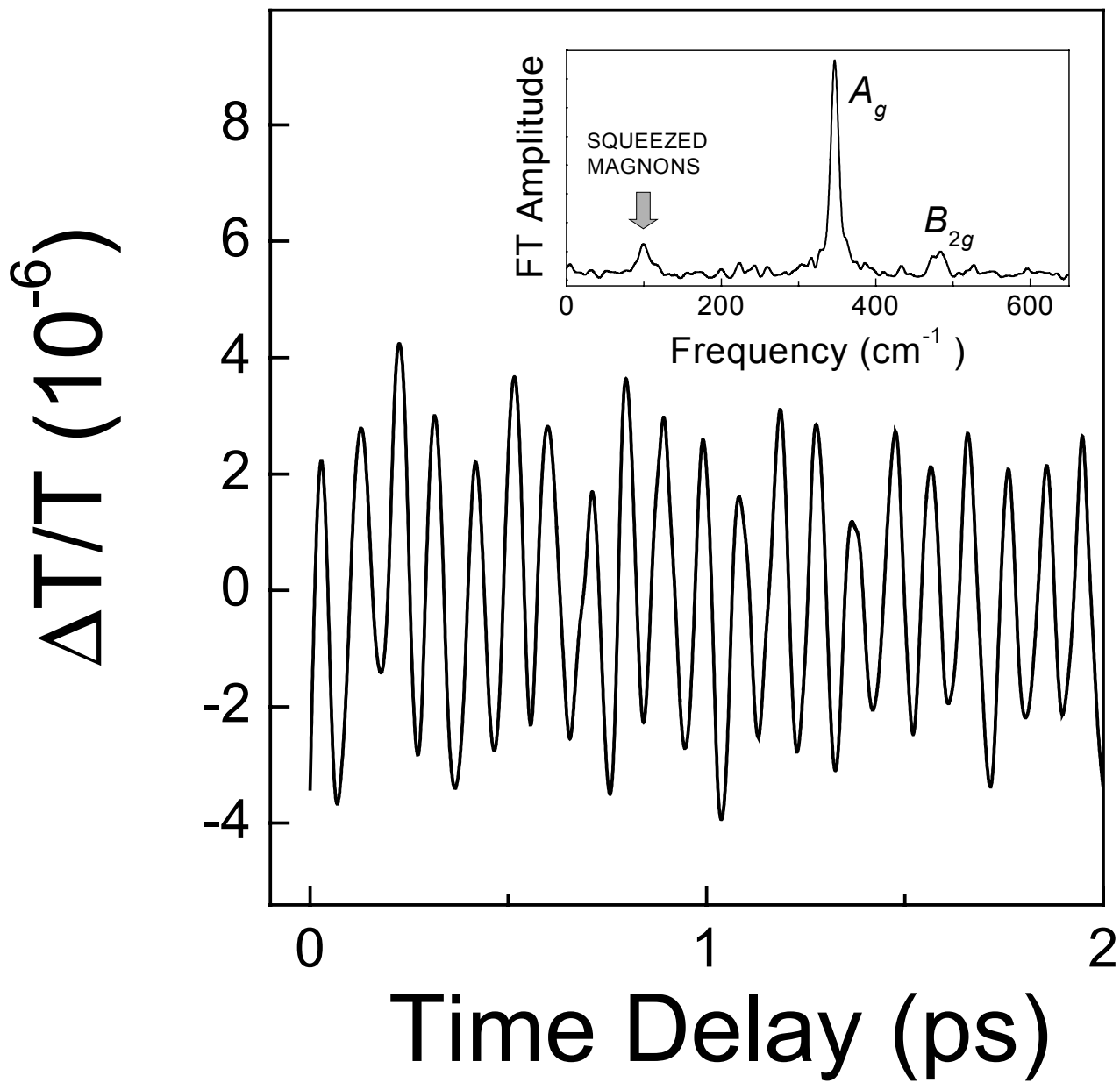

FIGURE 1

- 13 -

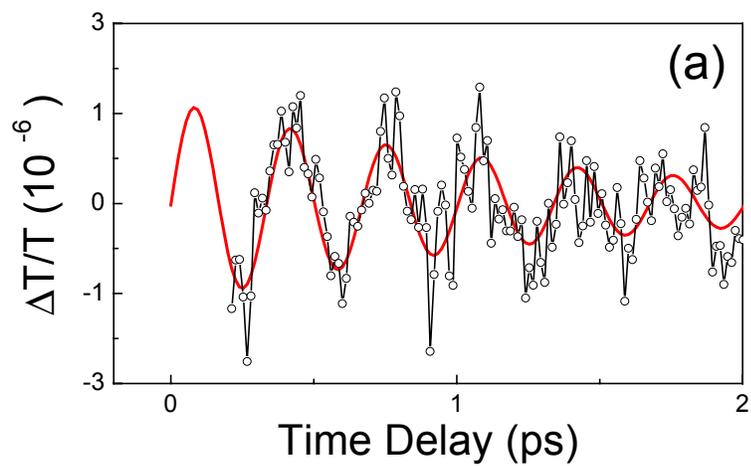

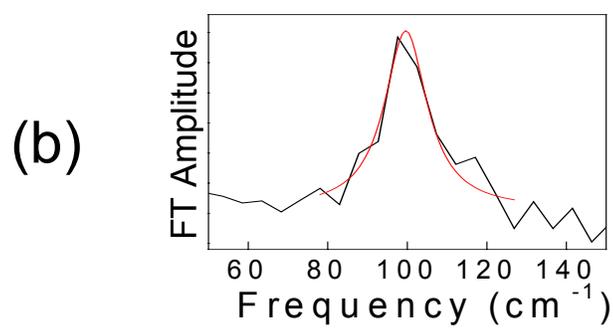

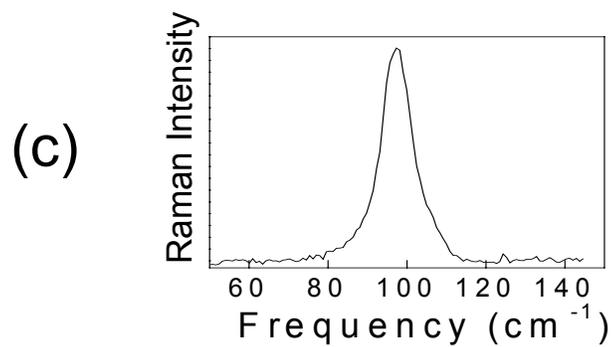

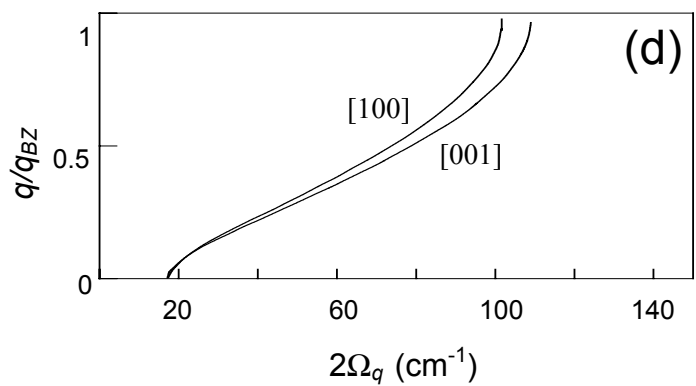

FIGURE 2

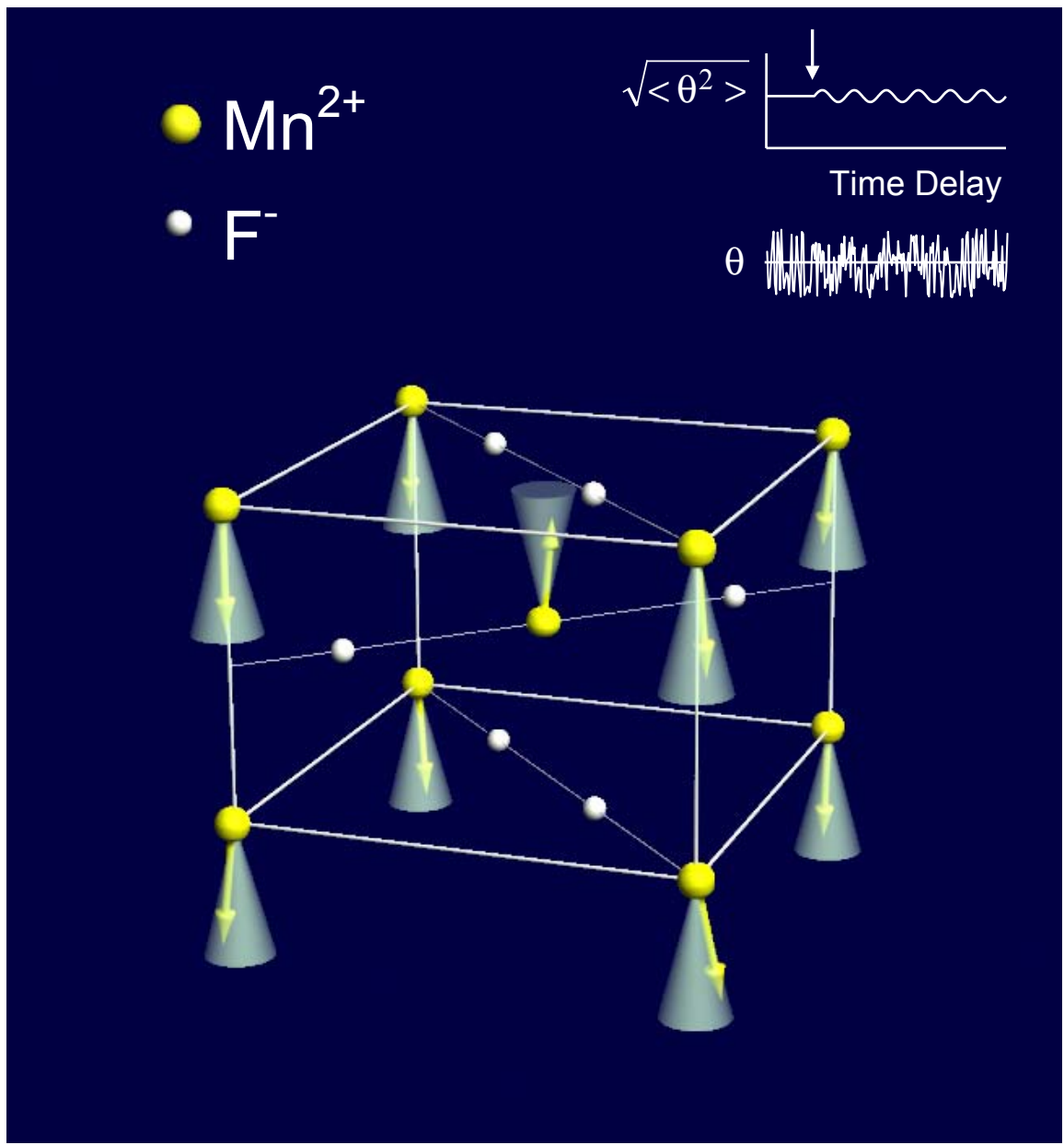

FIGURE 3